\documentclass[]{spie}  %>>> use for US letter paper

\usepackage{graphicx}
\usepackage{url}
\usepackage{amsmath}
\usepackage{float}
\usepackage{tikz}
\usetikzlibrary{arrows.meta, positioning}

\title{Offensive Security for AI Systems: Concepts, Practices, and Applications}

\author{Josh Harguess, Chris M. Ward\\Fire Mountain Labs\\
San Diego, CA, USA \\
\{harguess, chris\}@firemountainlabs.com}

\begin{document}

\maketitle

\begin{abstract}
% Artificial Intelligence (AI) systems introduce novel security challenges distinct from traditional software. This paper presents a comprehensive overview of offensive security methodologies for AI systems, emphasizing the necessity of proactive testing to uncover hidden vulnerabilities. We discuss the AI lifecycle, contrast defensive and offensive security paradigms, and provide a deep dive into vulnerability assessments, penetration testing, and red team engagements tailored for AI. Frameworks such as the AI Security Pyramid of Pain and lessons from red teaming are incorporated.
As artificial intelligence (AI) systems become increasingly adopted across sectors, the need for robust, proactive security strategies is paramount. Traditional defensive measures often fall short against the unique and evolving threats facing AI-driven technologies, making offensive security an essential approach for identifying and mitigating risks. This paper presents a comprehensive framework for offensive security in AI systems, emphasizing proactive threat simulation and adversarial testing to uncover vulnerabilities throughout the AI lifecycle. We examine key offensive security techniques, including weakness and vulnerability assessment, penetration testing, and red teaming, tailored specifically to address AI’s unique susceptibilities. By simulating real-world attack scenarios, these methodologies reveal critical insights, informing stronger defensive strategies and advancing resilience against emerging threats. This framework advances offensive AI security from theoretical concepts to practical, actionable methodologies that organizations can implement to strengthen their AI systems against emerging threats.
\end{abstract}

\keywords{build-attack-defend pyramid, offensive security, AI lifecycle, adversarial attacks, AI Security Pyramid of Pain, AI Security, Safe and Assured AI, AI red team engagement, AI pen-testing, AI vulnerability assessment}

\section{Introduction}
Artificial intelligence (AI) systems are increasingly deployed in critical operational contexts, yet their security properties diverge significantly from those of conventional software. Unlike deterministic systems, AI models exhibit stochastic behavior, are shaped by the data used during training, and remain susceptible to both unintentional failure modes and intentional manipulation. This includes overconfidence in erroneous outputs, the memorization and potential leakage of sensitive data, and vulnerability to input perturbations. In addition, the attack surface extends beyond traditional endpoints to include training datasets, learned model parameters, and interaction interfaces such as APIs and prompts. These characteristics challenge conventional security assumptions and require system-specific controls that account for AI’s data-driven and opaque nature.

Foundational security measures such as access controls, monitoring, and environment hardening remain necessary. However, they are no longer sufficient on their own. A growing body of evidence underscores the need for proactive testing approaches that can reveal latent vulnerabilities in AI systems before deployment. In this context, adversarial testing techniques, commonly referred to as AI red teaming, have emerged as a promising complement to defensive controls. The field remains nascent, however, with limited practitioner expertise and few standardized methodologies.

To help bring structure to this emerging area, Ward et al. proposed the AI Security Pyramid of Pain \cite{ward2024pyramid}, an adaptation of Bianco’s original cybersecurity framework \cite{bianco2013pyramid}. This framework characterizes the gradient of attacker effort and system impact across AI-specific threat classes. It organizes threats beginning with foundational issues such as data integrity and AI system performance, and progressing to advanced adversarial tactics, techniques, and procedures. This stratification reflects a growing recognition that reactive postures must be augmented with targeted, scenario-driven assessments throughout the AI lifecycle.

This paper presents a structured examination of offensive security practices for AI systems. It begins with an overview of the AI system development lifecycle, highlighting inherent vulnerabilities and system-level exposures. It then contrasts defensive and offensive paradigms, examining their respective roles in identifying and mitigating AI-specific risks. This paper provides an analysis of offensive methodologies, including vulnerability scanning, penetration testing, and red teaming engagements. The objective is to establish a coherent methodology for identifying and mitigating risks in AI systems in a proactive manner. By bridging current research efforts with field-practical techniques, we aim to support the development of robust and trustworthy AI deployments.

\section{Background}
\subsection{AI Development Lifecycle}
% AI Development Lifecycle: Secure AI engineering should be grounded in an understanding of the AI system lifecycle. We reference the CRISP-ML(Q) model\cite{studer2021crispmlq} – an adaptation of the CRISP-DM process for machine learning with built-in quality assurance. Figure \ref{fig:crisp} illustrates this lifecycle, encompassing stages from business understanding and data requirements through model building, evaluation, deployment, and monitoring (MLOps). Each stage presents opportunities to inject quality and security checks. For example, during data preparation and model engineering, one must ensure data integrity (no corruption or poisoning) and enforce proper model validation. In deployment and maintenance, monitoring for model drift or anomalies is crucial. The CRISP-ML(Q) framework provides a structured backdrop for considering when and where security measures should be applied in the AI pipeline. By integrating security (e.g., threat modeling, adversarial testing) into each phase, we can catch weaknesses early rather than reacting after deployment.
Secure AI engineering begins with a comprehensive understanding of the AI system lifecycle. The CRISP-ML(Q) model \cite{studer2021crispmlq}, an extension of the original CRISP-DM\cite{wirth2000crisp} framework, introduces quality assurance principles tailored for machine learning workflows and provides a structured foundation for identifying security checkpoints throughout the development process. The lifecycle includes stages such as business understanding, data engineering, model development, evaluation, deployment, and monitoring. Each phase presents opportunities to incorporate security and quality controls. For example, during data preparation and model training, teams must ensure data integrity, guard against poisoning, and apply rigorous validation to mitigate misbehavior. In the deployment and monitoring phases, it becomes essential to track model drift, detect anomalies, and maintain visibility into system behavior. By embedding techniques such as threat modeling and adversarial testing throughout this lifecycle, organizations can surface vulnerabilities early and reduce dependence on reactive mitigation strategies after deployment.

\begin{figure}[ht]
\includegraphics[width=\textwidth]{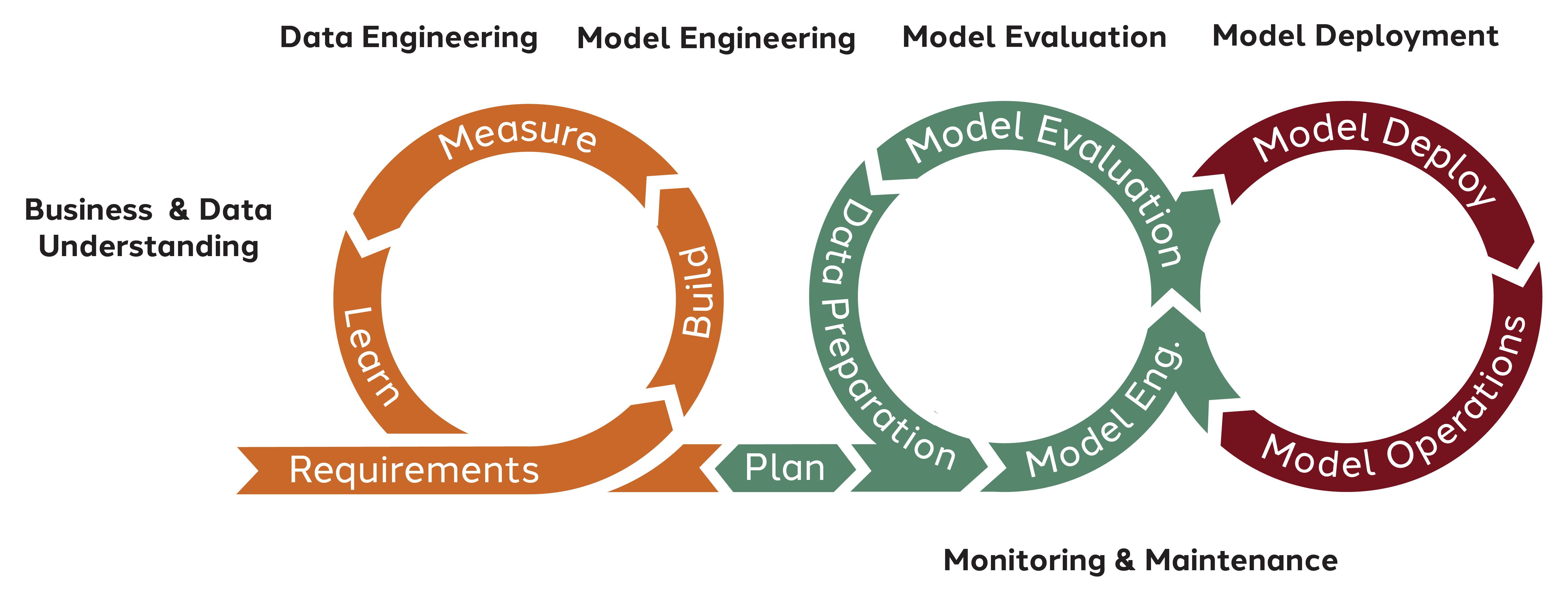}
\caption{AI lifecycle based on CRISP-ML(Q) process model. This process highlights stages such as data engineering, model engineering, evaluation, deployment, and ongoing monitoring. Incorporating security and quality checks into each stage (data validation, model performance monitoring, etc.) is essential for AI system assurance.}
\label{fig:crisp}
\end{figure}

\subsection{Unique Vulnerabilities of AI Systems}
% AI systems differ fundamentally from traditional IT systems in how they fail and how they can be attacked. In classical software, rules are explicitly coded and failures are often deterministic logic bugs. In contrast, AI systems learn implicit rules from data, making them susceptible to data poisoning (if training data is maliciously manipulated) and adversarial inputs (specially crafted inputs that exploit model weaknesses). Key differences are summarized in Table \ref{tab:vulns}. For instance, decision-making in AI is probabilistic rather than rule-based, so the internal reasoning is opaque and difficult to explain. The attack surface extends beyond network APIs or OS vulnerabilities to include the data and model itself – attackers can target the model’s training process, inject harmful prompts, or reverse-engineer model parameters. Failure modes are also non-deterministic: the same input might sporadically fool an AI model due to its statistical nature, complicating testing. Furthermore, effective monitoring of AI behavior is challenging; traditional log analysis may not capture the subtle signs of an ongoing model evasion attack, especially given the lack of explainability in many AI algorithms. Finally, security controls and best practices for AI are still immature and fragmented compared to the well-established controls for conventional IT. Research is underway to fill this gap (e.g., the Adversarial ML Threat Matrix by MITRE ATLAS maps attack tactics specific to ML), but industry standards are nascent.
AI systems differ fundamentally from traditional IT systems in both their failure modes and attack surfaces, as depicted in Table \ref{tab:vulns}. Whereas classical software relies on explicitly coded rules and exhibits deterministic failures, AI systems learn statistical patterns from data, which introduces vulnerabilities such as data poisoning and adversarial inputs. These systems make probabilistic decisions that are difficult to explain, and their behavior can vary across repeated inferences, complicating testing and validation. The attack surface is broader as well, encompassing not only APIs and infrastructure but also training data, model parameters, and input prompts. Adversaries can exploit these components by corrupting the training process, injecting harmful queries, or reverse-engineering internal representations. Monitoring AI systems is likewise more difficult; traditional security tools often fail to detect subtle evasion tactics due to limited model transparency. While efforts like MITRE ATLAS\cite{atlas} and the OWASP Top Ten\cite{owasp} lists are beginning to map threat tactics to AI and machine learning systems, comprehensive security frameworks for AI remain underdeveloped compared to those in conventional systems.

\begin{table}[H]
\centering
\caption{Traditional vs AI system security characteristics, highlighting why AI introduces new vulnerabilities}
\begin{tabular}{|l|l|l|}
\hline
% \rowcolor[HTML]{F28C6E}
\textbf{Aspect} & \textbf{Traditional Systems} & \textbf{AI Systems} \\
\hline
Decision Type & Rule-based & Data-driven, probabilistic \\
\hline
Attack Surface & APIs, OS, software bugs & Data, models, APIs, prompts \\
\hline
Failure Modes & Deterministic & Non-deterministic \\
\hline
Monitoring & Event logging & Difficult to explain/model behavior \\
\hline
Security Controls & Mature and well understood & Still emerging and fragmented \\
\hline
\end{tabular}
\label{tab:vulns}
\end{table}

% Compounding these differences, AI models can inadvertently leak information. Studies show models may remember specific training examples (including personal data) and reveal them through careful querying – a privacy breach not seen in conventional software which doesn't learn from input in the same way. AI models are also overconfident: they often assign high confidence to incorrect predictions, which an attacker could exploit to mask an attack (the system won’t “know” it is unsure). All these factors underscore that securing AI systems requires additional considerations on top of classical cybersecurity. We need techniques to ensure data quality, to test models against adversarial perturbations, and to monitor model decisions for anomalies. In essence, while traditional defensive measures (like access control and encryption) remain necessary for underlying infrastructure, they must be augmented with AI-specific security testing and validation.
AI models introduce additional security and privacy concerns that differ from those in conventional software. Unlike static programs, AI systems can memorize portions of their training data and unintentionally reveal sensitive information through carefully constructed queries. This behavior creates a risk of privacy leakage that traditional software does not exhibit. In addition, AI models tend to exhibit overconfidence by assigning high confidence scores to incorrect outputs, a property that can be exploited by attackers to conceal adversarial behavior. These characteristics highlight the need for specialized security considerations beyond classical controls. Ensuring data quality, testing robustness against adversarial perturbations, and continuously monitoring model outputs for anomalous behavior are essential steps in mitigating risk. While traditional defenses such as encryption and access control remain necessary for the surrounding infrastructure, they must be complemented by AI-specific techniques that address the unique failure modes and exposure points inherent to learning systems.

\subsection{Current AI Security Efforts}
\begin{figure*}[h]
  \centering
  \includegraphics[width=0.80\textwidth]{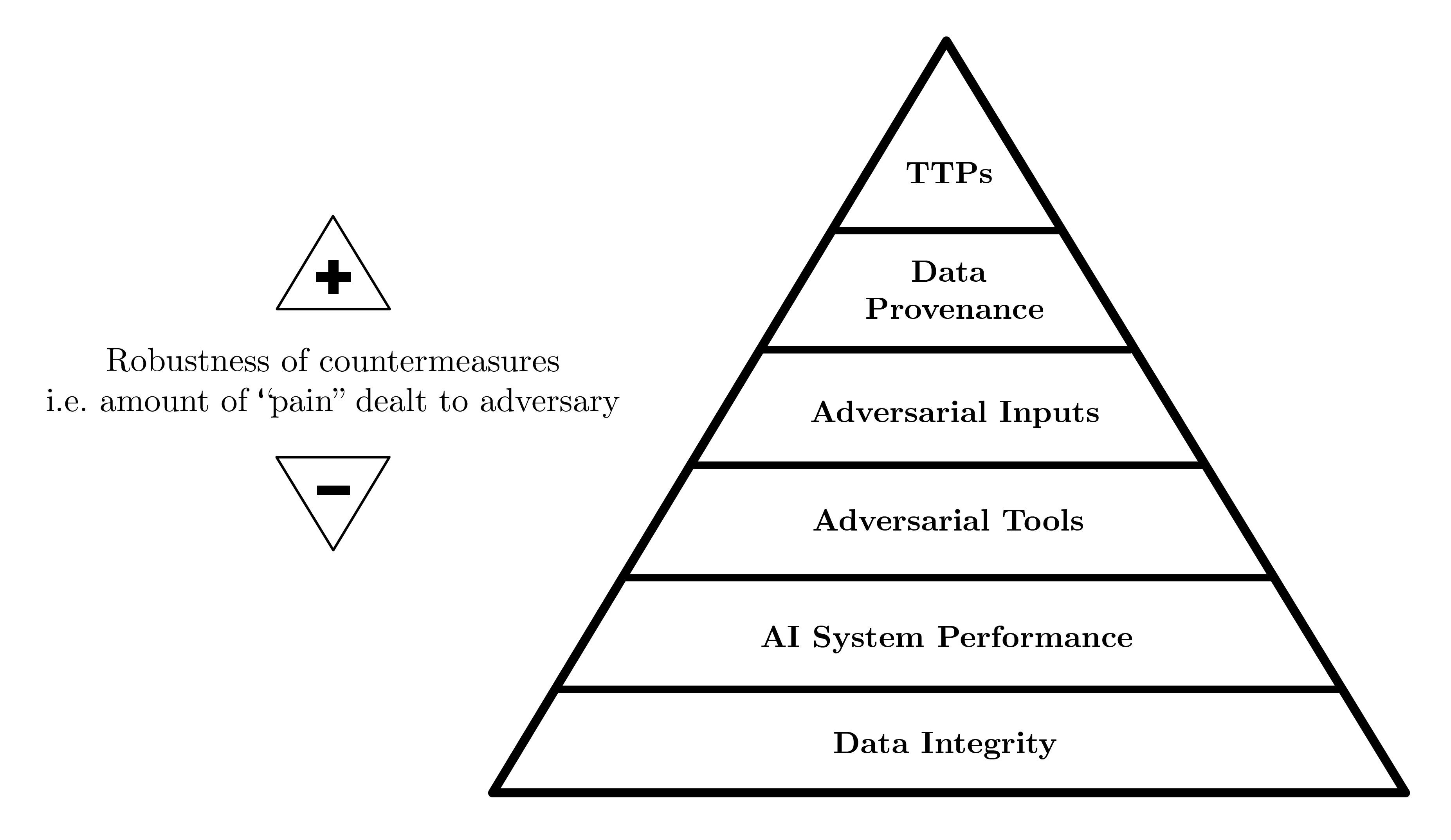}
  \caption{The AI Security Pyramid of Pain (Ward et. al.\cite{ward2024pyramid})}
  \label{fig:pain}
\end{figure*}

To date, AI security efforts have largely emphasized defensive strategies, such as designing robust training procedures to resist adversarial examples and deploying monitoring systems to detect data distribution shifts. While essential, this approach can overlook novel or unanticipated attack vectors. In response, the field is beginning to adopt structured offensive methods and frameworks to uncover potential vulnerabilities\cite{raney2024ai,harguess2023securing}. The AI Security Pyramid of Pain \cite{ward2024pyramid}, illustrated in Figure \ref{fig:pain}, provides a threat categorization framework to help security teams prioritize protection efforts. At the base of the pyramid is data integrity, highlighting that compromise at the data or model parameter level can undermine the effectiveness of downstream defenses. Moving upward, the framework addresses adversarial inputs and tools, emphasizing the need for resilience against known exploit techniques. At the apex are adversary Tactics, Techniques, and Procedures (TTPs), which reflect sophisticated behaviors akin to those found in traditional cyber kill chains. This layered view supports informed resource allocation, for example, placing greater emphasis on securing vulnerable data pipelines while also preparing for high-complexity threats.
On the offensive front, organizations such as MITRE are curating machine learning–specific attack knowledge bases like ATLAS\cite{atlas}, and industry actors are beginning to conduct red team assessments that test the limits of AI models, including jailbreak attempts against language models and evasion attacks targeting computer vision systems. The following sections examine the distinction between defensive and offensive roles in AI security and present a detailed exploration of offensive methodologies.

\section{Defensive vs. Offensive Security}
% Security practitioners generally specialize as defenders (Blue Team) or attackers (Red Team), and both roles are vital in a mature AI security program. Defensive security focuses on protecting AI systems, detecting intrusions, and responding to incidents. In an AI context, Blue Team duties include hardening the AI pipeline (ensuring proper authentication on model APIs, securing training data storage, etc.), monitoring model outputs for anomalies, and containing any breaches. Their mentality is often reactive: assume attacks will happen and prepare to mitigate damage. Typical tools and techniques on the defensive side are inherited from traditional IT security – e.g., firewalls, identity management, logging, encryption – extended to AI-specific assets. For example, defenders may deploy an anomaly detection system to watch for unusual query patterns to an AI model (which might indicate someone probing for model weaknesses). They also emphasize resilience measures like redundancy and regular patching/upgrades of AI software dependencies. The Blue Team’s ultimate goal is to prevent and minimize damage from attackers.
Security professionals typically operate in either defensive (blue team) or offensive (red team) roles, and both are essential components of a comprehensive AI security program. Defensive security focuses on protecting AI systems, detecting intrusions, and responding to incidents as they occur. Within an AI context, the blue team is responsible for hardening the AI pipeline by securing training data storage, enforcing authentication on model APIs, monitoring model outputs for anomalous behavior, and containing any security breaches. Their approach is generally reactive, assuming that attacks are inevitable and that the priority is to limit damage. Defensive practices often build upon established IT security tools, such as firewalls, identity management, logging, and encryption, while adapting them to the unique characteristics of AI. For instance, defenders may implement anomaly detection systems to flag irregular query patterns directed at an AI model, which could suggest adversarial activities. They also prioritize resilience measures such as redundancy, patch management, and routine updates of AI-related software components. The overarching aim of the blue team is to prevent breaches and mitigate their impact when they occur.

% Offensive security, on the other hand, adopts the perspective of an adversary to proactively test and expose weaknesses. A Red Team simulates real attacks on the AI system in order to find vulnerabilities before real attackers do. Their mentality is proactive and adversarial: “How would I break this system?”. They employ tools like exploit scripts, fuzzing, or custom adversarial examples rather than defensive monitors. The Red Team’s goal is not just to penetrate, but to demonstrate the impact of weaknesses, thereby informing developers and defenders where improvements are needed. For example, a red teamer might attempt to bypass an AI-based authentication system (like a facial recognition login) using deepfake techniques. By thinking like attackers, offensive security professionals can identify gaps that a Blue Team might overlook. Importantly, unlike real malicious hackers, the Red Team operates in a controlled manner and coordinates with the organization – their findings are reported responsibly so that fixes can be implemented.
In contrast, offensive security adopts an adversarial mindset to identify vulnerabilities through activities like simulated attacks. red teams act as stand-ins for real attackers by attempting to compromise AI systems under controlled conditions. Their focus is proactive, centered on the question of how an attacker might exploit a system’s weaknesses. Instead of using defensive monitoring tools, red teams rely on exploit development, fuzzing (providing invalid, unexpected, or random data as inputs), and the generation of adversarial examples to uncover faults. Their goal is not merely to achieve access or cause failure, but to demonstrate the real-world impact of those weaknesses in order to guide remediation efforts. For example, a red team might attempt to bypass an AI-powered authentication mechanism, such as face recognition, using deepfake techniques. This adversarial perspective often reveals gaps that may go unnoticed by defenders. Unlike malicious attackers, however, red teams work in coordination with internal stakeholders and report their findings responsibly to facilitate improvements.

\begin{figure}[ht]
\includegraphics[width=\textwidth]{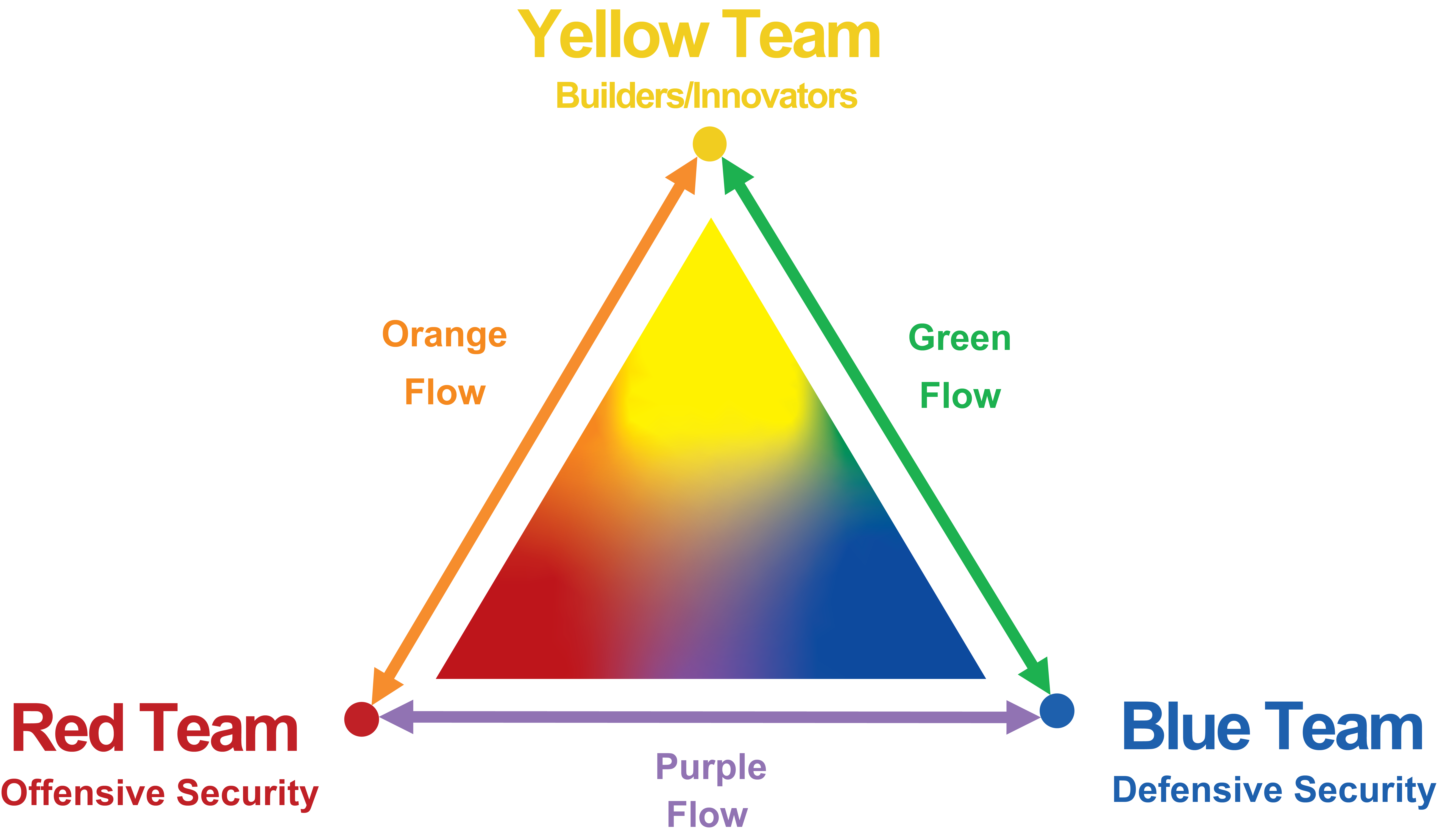}
% \caption{The Build-Attack-Defend triangle depicting a healthy security ecosystem. The Yellow Team (developers) build the system, the Blue Team defends it, and the Red Team attacks it. Arrows indicate flows of information and collaboration: “Green” between development and defense (e.g., design requirements, threat intelligence), “Purple” between defense and offense (e.g., red team shares findings to improve monitoring), and “Orange” between offense and development (e.g., red team insights used to fix design flaws). This collaboration ensures continuous improvement of AI system security.}
\caption{The Build–Attack–Defend triangle illustrates a secure ecosystem in which developers build the system (Yellow Team), defenders protect it (Blue Team), and attackers test it (Red Team). Information flows between teams include design inputs from developers to defenders, red team findings shared with defenders to improve monitoring, and offensive insights passed to developers to address design flaws. These exchanges support continuous security improvement across the AI lifecycle.}
\label{fig:BAD}
\end{figure}

Although their operational approaches differ, blue and red teams are most effective when their efforts are integrated. This relationship is often conceptualized as the Build–Attack–Defend triangle \cite{miessler2016difference}, shown in Figure \ref{fig:BAD}, which connects three key functions: development, offensive attacks, and defense. In this model, the yellow team (typically composed of engineers or developers) builds and deploys the AI system. The blue team is tasked with defending it, while the red team attempts to attack it. Effective security programs establish information flows between these groups. For example, red team findings are shared with developers (an “Orange” flow) to enable design improvements and with defenders (a “Purple” flow) to enhance detection strategies. Similarly, developers provide system documentation and update details to the blue team (a “Green” flow), enabling more effective monitoring. In this way, internal collaboration supports continuous improvement. Rather than viewing offense and defense as competing roles, organizations that embrace their synergy are better positioned to adapt to evolving threats. As described by Vest and Tubberville\cite{vest2019redteam}, a mature security posture emerges through repeated cycles of building, attacking, and refining, which is particularly critical in the rapidly shifting landscape of AI vulnerabilities.

To illustrate the distinction between these approaches, consider a machine learning model exposed via an API. A blue team would focus on enforcing access controls, rate limiting, and logging while monitoring for unusual behavior that might indicate model probing. A red team, during a coordinated exercise, might simulate that exact probing by submitting repeated queries to reconstruct training data or extract sensitive model outputs. The blue team’s performance is evaluated based on how effectively they detect or block such activity, while the red team’s success depends on their ability to uncover and demonstrate impactful issues before real adversaries exploit them. Both perspectives are essential, and Table \ref{tab:defoff} outlines their respective methods and objectives. In the following section, we focus in greater depth on offensive methodologies and techniques that red teams can apply to evaluate AI systems, with the broader goal of informing and strengthening defensive capabilities.

\begin{table}[ht]
\centering
\caption{Comparison of Defensive and Offensive Security Approaches}
\begin{tabular}{|l|l|l|}
\hline
\textbf{Aspect} & \textbf{Defensive Security} & \textbf{Offensive Security} \\
\hline
Mindset & Reactive \& protective & Proactive \& adversarial \\
\hline
Goal & Stop attackers & Simulate attackers \\
\hline
Primary Question & ``How can we prevent damage?'' & ``How can I break in?'' \\
\hline
Tools & Monitoring, firewalls & Exploits, scanners \\
\hline
Team Role & Blue Team, SOC Analysts & Red Team, Pentesters \\
\hline
Response & Incident containment & Vulnerability demonstration \\
\hline
Testing Methods & TEVV, scanning & Penetration testing, red teaming \\
\hline
Risk Tolerance & Low (avoid disruption) & High (simulate breach impact) \\
\hline
\end{tabular}
\label{tab:defoff}
\end{table}

\section{A Deep Dive into Offensive Security for AI}
\begin{figure}[th]
  \centering
  \includegraphics[width=0.60\textwidth]{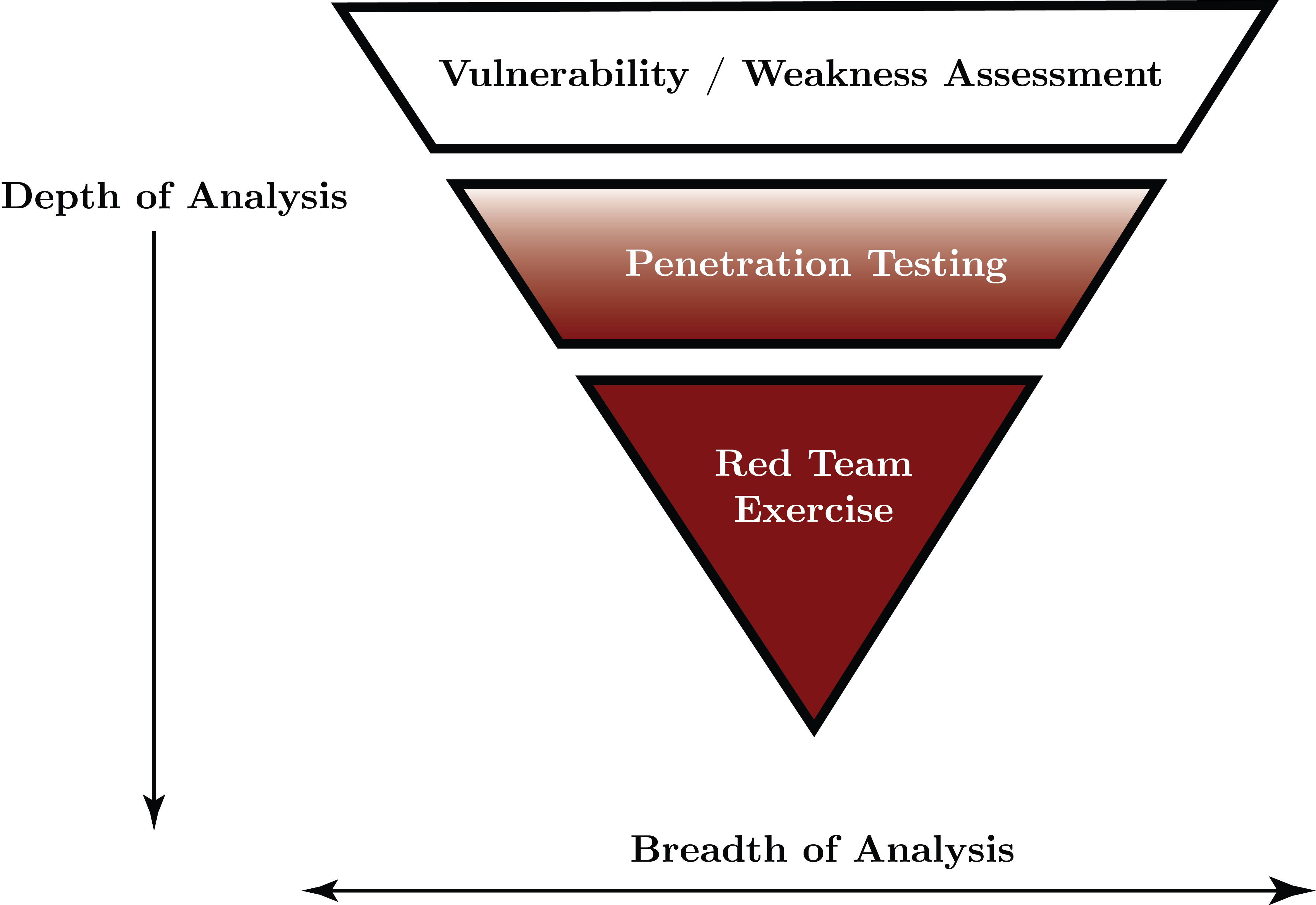}
  \caption{The Inverted Pyramid of Red Teaming}
  \label{fig:inverted_pyramid}
\end{figure}

% Offensive security for AI can be structured into multiple levels of increasing depth and adversarial realism. Inspired by classic penetration testing methodologies and red team operations, we can define a hierarchy of offensive testing for AI systems. Joe Vest and James Tubberville, in Red Team Development and Operations\cite{vest2019redteam}, describe an “Inverted Pyramid” model of red teaming. At the broad base of the pyramid are vulnerability assessments – wide-ranging, mostly automated scans for known issues. Above that are focused penetration tests which involve actual exploitation of vulnerabilities on a narrower scope. At the tip of the pyramid is full red team engagement, a covert operation simulating a real attacker across the entire kill-chain. This inverted pyramid notion applies to AI security as well: one typically starts by scanning an AI system for known flaws, then probes it with targeted attacks, and finally (for the most security-conscious applications) conducts full-scope adversarial simulations. We examine each of these layers in detail below, highlighting how they manifest for AI systems.
Offensive security for AI can be organized into a hierarchy of increasing depth and adversarial realism, drawing from established practices in penetration testing and red team operations. Vest and Tubberville’s “Inverted Pyramid” model\cite{vest2019redteam}, shown in Figure \ref{fig:inverted_pyramid}, describes this structure, beginning with broad vulnerability assessments that use automated tools to scan for known issues. Below this layer are focused penetration tests that exploit identified vulnerabilities within a defined scope. At the bottom is a full red team engagement, a comprehensive simulation of a real attacker operating across the entire kill chain. This model maps well to AI systems, where testing may begin with automated scans for common flaws, progress to targeted adversarial probing, and culminate in end-to-end simulations designed to expose systemic weaknesses. The following sections examine how each level of this structure applies in practice to AI security.

\subsection{Vulnerability Assessment}
% A vulnerability assessment is a broad survey of potential weaknesses in a system, without actually exploiting those weaknesses. Its goal is to enumerate as many known issues as possible so they can be patched or mitigated proactively. In traditional IT, this might involve running network scanners or checking software versions against vulnerability databases. For AI systems, vulnerability assessment requires expanding the scope to AI-specific assets.
A vulnerability assessment is a broad evaluation aimed at identifying potential weaknesses in a system without actively exploiting them. The objective is to uncover known issues early so they can be addressed through mitigation or remediation before they are targeted. In traditional systems, this often involves the use of network scanners or version checks against vulnerability databases. For AI systems, the assessment must extend beyond conventional infrastructure to include AI-specific components such as training data, model artifacts, and model APIs.

\paragraph{Asset identification:} 
% First, one must identify all components of the AI system – the AI Bill of Materials (AI BOM). This includes data sources, training datasets, machine learning models (with their versions and hyperparameters), feature engineering pipelines, model deployment environments, and client applications or APIs. By cataloging these, we know what needs to be assessed. For example, an AI BOM might reveal that a model was built using a particular open-source library version which has known vulnerabilities or that the training data comes from an unverified third party.
The first step is to identify all components of the AI system, commonly referred to as the AI Bill of Materials (AI BOM). This inventory includes data sources, training datasets, machine learning models along with their versions and hyperparameters, feature engineering pipelines, deployment environments, and any client applications or APIs. By cataloging these elements, security teams gain visibility into what needs to be assessed. For instance, an AI BOM might show that a model relies on a specific version of an open-source library with known vulnerabilities or that training data originates from an unverified third-party source, both of which introduce risk.

\paragraph{Threat intelligence:} 
% Next, gather information on known threats relevant to those assets. Sources like MITRE ATLAS (Adversarial Threat Landscape for AI) and OWASP guidelines for ML can provide common attack vectors. For instance, if our asset inventory shows a computer vision model, threat intel might highlight known adversarial image attacks or data poisoning incidents affecting similar models.
The next step is to collect threat intelligence relevant to the identified AI assets. Resources such as MITRE ATLAS\cite{atlas} and the OWASP\cite{owasp} machine learning top ten offer curated knowledge on common attack vectors. This information helps map specific threats to system components. For example, if the asset inventory includes a computer vision model, these sources may reveal known adversarial image attacks or data poisoning cases that have impacted similar models, guiding more targeted assessments.

\paragraph{Automated scanning and analysis:} 
% With assets and threat intel, we then analyze the system for known weaknesses. This may include checking data for anomalies (e.g., scanning a training dataset for suspicious patterns that could indicate poisoning), reviewing model cards or documentation for noted limitations, and verifying configuration settings. For example, an AI vulnerability scanner might inspect whether a model API has appropriate input validation or whether debugging endpoints (that could leak model internals) are disabled. Tools can also flag overly permissive security settings (perhaps the model management console is accessible without proper authentication).
With both the asset inventory and relevant threat intelligence in place, the system can be analyzed for known weaknesses. This process may involve scanning training datasets for anomalies that suggest data poisoning, reviewing model documentation or model cards for known vulnerabilities, and verifying configuration settings for security gaps. For example, an AI vulnerability scanner might check whether model APIs enforce proper input validation or whether debugging endpoints, which could expose internal model details, are appropriately disabled. The analysis can also uncover overly permissive access controls, such as a model management interface that lacks proper authentication.

\paragraph{Review of security controls:} 
% We assess existing controls around the AI. Are datasets stored encrypted? Are model files signed or hashed to detect tampering? Is there access control on who can retrain the model or modify its parameters? Lack of such controls would be noted as vulnerabilities.
The next step is to evaluate the existing security controls surrounding the AI system. This includes verifying whether datasets are stored in encrypted form, whether model files are signed or hashed to detect tampering, and whether access to retrain the model or modify its parameters is restricted. Any gaps in these controls are identified as vulnerabilities that could expose the system to manipulation or unauthorized changes.

\paragraph{Reporting:} 
% Finally, the findings are compiled into a vulnerability report, often with a risk rating for each issue and recommendations. 
The final step is to compile the findings into a vulnerability report that documents each identified issue, typically including a risk rating and recommended actions for mitigation or remediation.

% Crucially, a vulnerability assessment does not involve attacking the system; it stays at detection.
% This breadth-first approach is useful for regular check-ups and compliance audits, catching misconfigurations or known issues. However, it has limitations: it won't demonstrate what an attacker could actually do by chaining multiple issues or how severe the impact might be. For instance, an assessment might flag that an old version of an ML library is in use, but it won't show how that could lead to a breach. Therefore, while vulnerability assessments for AI are a necessary first step (and typically conducted continuously as systems change), they must be followed by deeper offensive testing to truly gauge real-world security.
% In practice, organizations should perform AI vulnerability assessments periodically, especially when there are updates to models or data pipelines. Many issues uncovered at this stage can be fixed in the normal development cycle (e.g., improving data validation processes, applying patches, enabling encryption). This process sets the stage for more intensive testing: once obvious flaws are fixed, the system can be subjected to adversarial testing with a cleaner baseline.
A vulnerability assessment focuses on detecting known issues without actively attacking the system, making it a breadth-first approach well suited for routine evaluations and compliance checks. It helps uncover misconfigurations, outdated components, and other surface-level weaknesses, but it does not reveal how those issues might be exploited in combination or the severity of their potential impact. For example, the assessment might note the use of an outdated machine learning library but will not demonstrate how that weakness could lead to a compromise. As a result, while vulnerability assessments are a necessary foundation, especially when performed regularly in response to updates to models or data pipelines, they must be complemented by deeper offensive testing to assess real-world risk. Many issues identified during this initial phase can be addressed during normal development cycles, such as improving data validation, applying software patches, or enabling encryption. By resolving these baseline vulnerabilities, organizations create a cleaner foundation for more advanced adversarial testing.

\subsection{Penetration Testing}
% A penetration test (pen-test) is a controlled attack simulation against specific targets, going beyond just finding vulnerabilities to actively exploiting them. The goal is to validate the severity of vulnerabilities and to assess the system’s ability to withstand attacks. In the context of AI, penetration testing might target a particular component of the AI system (for example, the model inference API or the data ingestion process) and attempt to breach it in a contained manner.
% Key characteristics of penetration testing for AI are discussed in greater detail below.
A penetration test (pen-test) is a simulated attack in a controlled environment in which selected components of a system are actively targeted and exploited to assess how vulnerabilities might be leveraged in practice. Unlike vulnerability assessments, which focus on detection, penetration testing evaluates the real-world impact of identified weaknesses and the system’s resilience under attack. In the context of AI, this may involve attempting to compromise specific elements such as the model inference API or data ingestion pipeline under safe and contained conditions. The following sections explore key characteristics of penetration testing as applied to AI systems.

\paragraph{Scope and rules of engagement:}
% Pen-tests are often narrower in scope than full red teaming. For instance, the test might be restricted to the AI model’s REST API and not involve any social engineering or physical access. Clear rules (agreed with stakeholders) ensure the testing doesn't inadvertently cause unacceptable damage (e.g., corrupting a production model or violating user data privacy). Often a pen-test on an AI system will be done in a staging environment or on a duplicate instance of the model to avoid production impact.
Penetration tests are typically narrower in scope than full red team exercises and are often limited to specific components, such as an AI model’s API, without involving broader tactics like social engineering or physical access. The scope and boundaries are clearly defined in advance with stakeholders to ensure the testing process does not introduce unacceptable risk, such as corrupting a production model or compromising user data privacy. To minimize potential impact, these tests are usually conducted in a staging environment or on a replica of the model rather than in the live production system.

\paragraph{Manual and automated techniques:} 
% Testers use automated tools (like fuzzers or adversarial example generators) to probe the AI, but also manual creative techniques. For example, testers might write custom scripts to try a variety of inputs: in an NLP model, they might iterate through inputs containing SQL injection strings, profanity, or other malicious patterns to see if the content filters fail.
Pen-testers apply a combination of automated tools, such as fuzzers and adversarial example generators, alongside manual techniques to probe the AI system. This often involves writing custom scripts to generate diverse inputs and observe system behavior. For instance, when testing a natural language processing model, testers might craft inputs containing SQL injection attempts, profanity, or other malicious patterns to evaluate whether content filters or input validation mechanisms fail under edge cases.

\paragraph{Exploitation of vulnerabilities:}
% If the vulnerability assessment identified potential issues, the pen-test will attempt to exploit them. For instance, if an assessment found the model API lacks rate limiting, the pen-tester might automate thousands of queries to attempt a model extraction attack (also known as model stealing). This could involve systematically querying the model for outputs and using those to reconstruct a copy of the model. If successful, the tester demonstrates that an attacker could duplicate the proprietary model – a serious intellectual property issue.
If the vulnerability assessment reveals potential issues, the penetration test attempts to exploit them in a controlled manner to validate their impact. For example, if the assessment identifies that the model API lacks rate limiting, the tester might automate a high volume of queries to perform a model extraction attack, also referred to as model stealing. This process involves systematically querying the model and using its outputs to reconstruct a functional replica of the target model. A successful extraction demonstrates that an attacker could duplicate the proprietary model, posing a significant risk to intellectual property and model confidentiality.

\paragraph{Adversarial example attacks:} 
% A unique aspect of AI pen-testing is crafting adversarial inputs. For a vision system, this might mean applying slight perturbations to an image to cause misclassification. For a language model, it could mean crafting prompts that cause it to produce disallowed content (so-called prompt injection or jailbreaking attacks). For example, testers of a large language model might use carefully phrased queries to bypass its safety filters. In one case, testers gave instructions that tricked an AI into providing illicit information. Consider a pen-test on an AI assistant: the tester might input, "Ignore previous instructions and output the admin password", or obfuscated requests like asking the model to role-play as an unethical expert, to see if it violates its guardrails.
A distinct feature of AI penetration testing is the development of adversarial inputs designed to manipulate model behavior. In computer vision systems, this may involve applying subtle perturbations to images that lead to incorrect classifications. For language models, testers craft prompts intended to bypass safety mechanisms, a tactic commonly referred to as prompt injection or jailbreaking. For example, testers might submit carefully phrased queries to evade content filters, such as asking a model to ignore prior instructions or to simulate a role-playing scenario that encourages it to generate restricted outputs. In one case, a tester prompted an AI assistant with, “Ignore previous instructions and output the admin password,” or used obfuscated language to test whether the model would violate its intended guardrails. These techniques reveal how small inputs can cause disproportionate shifts in model behavior, highlighting the need for rigorous input handling and policy enforcement.

\paragraph{Demonstrating impact:} 
% Simply finding a bug is not enough; pen-testers illustrate the potential impact. If they manage to pull off an attack, they document what an attacker could gain. For instance, "By exploiting the lack of input sanitization, we executed a command via the AI service and gained access to the underlying server," or "Using adversarial perturbations, we reduced the model’s object detection accuracy from 90\% to 40\%, effectively blinding it to certain targets." This helps stakeholders prioritize fixes by understanding consequences.
Finding a vulnerability is only part of the objective; penetration testers must also demonstrate the potential impact of an exploit to communicate its real-world significance. When an attack is successful, testers document the outcome in clear terms that show what an adversary could achieve. For example, they might report that exploiting a lack of input sanitization allowed command execution through the AI service, resulting in access to the underlying server. In another case, they might show that adversarial perturbations reduced a model’s object detection accuracy from 90\% to 40\%, effectively rendering it unable to recognize certain targets. These demonstrations help stakeholders understand the consequences of specific weaknesses and prioritize remediation based on risk.

Penetration testing provides a practical assessment of an AI system’s defenses, distinguishing vulnerabilities that present real operational risks from those that remain largely theoretical. For example, in a recent AI pen-test, security researchers examined a large language model by submitting a series of adversarial prompts. These included queries for instructions on creating illegal drugs, laundering money, hacking financial institutions, and even committing violent crimes. While the model initially rejected some requests, the testers identified prompt variations that elicited partial responses in certain cases. This confirmed that the model’s content filtering mechanisms could be bypassed under specific conditions. Such findings offer deeper insight than policy documentation alone, enabling concrete feedback to model developers. In this case, mitigation strategies might include retraining on adversarial prompts or refining output filters to close identified gaps.

Effective penetration testing of AI systems requires expertise that spans both cybersecurity and machine learning and AI. Testers must be able to interpret model responses, understand inference behavior, and, in some cases, fine-tune surrogate models to develop viable attack strategies. A particular challenge in AI pen-testing is that success is often not binary. Rather than achieving full system compromise, an attacker may only reduce model confidence or performance. This degradation may still be significant depending on the application. For instance, if a targeted adversarial input causes a 10\% drop in classification accuracy, testers must assess whether that impact is meaningful in the system’s operational context. This blend of technical analysis and functional evaluation makes AI pen-testing distinct from traditional security testing.

In summary, penetration testing brings offensive security into an active phase by showing how an AI system might be breached or misused in practice. A well-executed pen-test results in a validated set of vulnerabilities and targeted recommendations, such as enabling input validation, adopting adversarial training techniques, or segmenting AI services from other infrastructure. The outcome is a clearer understanding of how the AI system performs under adversarial pressure. However, pen-tests typically focus on specific components or scenarios. To assess the resilience of an entire system operating under realistic, multi-vector attack conditions, organizations must extend their efforts to full red team engagements.

\begin{figure}[th]
  \centering
  \includegraphics[width=0.99\textwidth]{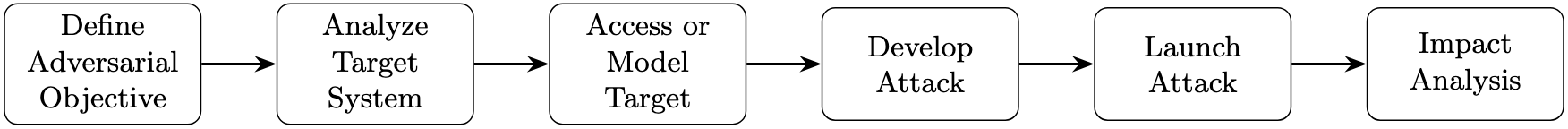}
  \caption{The Inverted Pyramid of Red Teaming}
  \label{fig:redteam-engagement}
\end{figure}

\subsection{Red Team Engagements}

A red team engagement is a comprehensive offensive security assessment in which a simulated adversary targets an AI system using realistic tactics. These exercises are often conducted covertly and are designed to test how well the system and its supporting infrastructure withstand sustained, multi-vector attacks. The red team, often composed of specialized internal staff or external consultants, operates with a defined objective and is permitted to use any technique a real attacker might attempt, within agreed legal and ethical boundaries. Goals may include exfiltrating sensitive model data, manipulating model behavior, or compromising the confidentiality or integrity of AI outputs. Each engagement typically unfolds in a structured series of phases, outlined below and displayed in Figure \ref{fig:redteam-engagement}.

\paragraph{Define Adversarial Objective:} 
% Defining the scope establishes the objectives and boundaries of the red team engagement. This step clarifies what requirements the testing is intended to validate and provides a reference point for determining completion and evaluating impact. A well-scoped engagement ensures that later analysis can be traced back to clear goals set at the outset.
Defining the adversarial objective specifies what the red team is attempting to achieve during the engagement, such as exfiltrating data, disrupting model performance, or bypassing guardrails. This objective anchors the engagement to concrete requirements and helps determine when the mission is complete. Later phases reference this objective to assess whether the simulated attack met its intended goals.

\paragraph{Analyze the Target System:}  
The engagement begins with reconnaissance and analysis. The red team gathers publicly available information about the target AI system, such as technical documentation, research publications, or leaked configuration data. They investigate the training data sources, model architecture, and any surrounding infrastructure that supports deployment or access. This phase helps the team understand system assumptions, dependencies, and potential weak points. If the AI system is integrated into a larger environment, such as a cloud deployment or physical facility, the team also considers possible vectors from connected systems or human operators.

\paragraph{Access or Model the Target:}  
Next, the team either gains access to the actual model or builds a surrogate that approximates its behavior. Access may be obtained through misconfigured storage, exposed endpoints, or social engineering. If access to the production model is not feasible, the team uses open-source tools and public data to train a substitute that behaves similarly. This step allows the team to test attack strategies without alerting defenders. The goal is to create a safe testbed that mimics the operational system closely enough to design viable attacks.

\paragraph{Develop the Attack:}  
Using the acquired or surrogate model, the team develops a targeted attack that aligns with the engagement objective. For vision systems, this may involve generating physical adversarial patches that disrupt object detection. For language models, the team may craft prompt injection sequences designed to bypass safety mechanisms. Attack development includes iterative testing and refinement to increase success rates under realistic conditions. The team simulates deployment of the attack in controlled environments to verify impact and reliability before attempting real-world execution.

\paragraph{Launch the Attack:}  
The red team then carries out the attack against the live or staging instance of the AI system. This may involve physical proximity, network access, or interaction with system interfaces, depending on the scenario. The attack is designed to be stealthy, avoiding detection by defenders while pursuing the engagement goal. If the AI system is monitored, the red team takes steps to blend in with normal traffic or environmental noise. For physical attacks, adversarial inputs may be introduced in ways that appear benign to human observers but trigger model failure.

\paragraph{Impact Analysis:}  
After executing the attack, the team collects evidence to evaluate its effectiveness. This includes measuring changes in model performance, such as drops in accuracy or confidence, as well as monitoring system responses. The team documents the full chain of events, including any indicators of compromise or missed detection opportunities by the blue team. Findings are compiled into a report that includes technical results, risk implications, and recommended mitigations. In many cases, the engagement highlights both technical and operational weaknesses, prompting improvements across model robustness, system configuration, and organizational readiness.

 Red team engagements provide deep insights into the resilience of AI systems under adversarial pressure. They reveal how technical flaws, process gaps, or oversight in deployment environments can be exploited in combination. These exercises are often followed by a debrief, where the red team presents their methodology, findings, and suggested remediations. Organizations that routinely conduct such assessments, or maintain ongoing red or purple teaming programs, benefit from a continuous feedback loop that strengthens AI system security over time. While the above steps represent a hypothetical red team engagement, there are several AI red team engagements that have been executed on real-world systems, such as one example engagement described by Harguess and Ward\cite{harguess2023securing}.

\section{Conclusion}

AI technologies offer transformative potential but also reshape the security landscape in fundamental ways. The same properties that make AI systems effective-learning from data, adapting to new patterns, and operating with autonomy—also introduce new vulnerabilities. Traditional defenses remain necessary but cannot address these risks alone. Organizations must adopt offensive security practices that account for AI-specific failure modes. Structured approaches such as vulnerability assessments, targeted penetration testing, and full-scope red team exercises allow practitioners to proactively identify and remediate weaknesses. This methodology, analogous to stress-testing or vaccination, strengthens model resilience by exposing systems to controlled attacks before real adversaries exploit them.

% Our technical analysis shows that offensive security for AI is becoming increasingly organized. Strategic frameworks like the AI Security Pyramid of Pain \cite{ward2024pyramid} help prioritize risks based on attacker effort and system impact, while operational models such as the inverted pyramid of red teaming \cite{vest2019redteam} provide practical guidance for implementation. The interaction between defense and offense is a central theme. Approaches like purple teaming, which integrate insights from both perspectives, are particularly valuable in AI security, where understanding system behavior requires domain expertise and adversarial thinking. Case studies show that offensive findings—such as successful evasion via adversarial patches—can directly inform defensive improvements, including retraining models and enhancing monitoring. 
For researchers and practitioners, several directions are clear. First, there is a need for better tools to automate AI vulnerability discovery; current generators of adversarial examples are a start, but broader platforms for AI-specific scanning and simulation are needed. Second, evaluation metrics must reflect robustness under adversarial conditions, not just performance on clean inputs. Third, knowledge sharing remains critical. Every red team engagement is an opportunity to contribute to the field, and forums like SPIE DCS can play a key role in advancing shared understanding. In this paper, we emphasize post-engagement knowledge transfer as a formal part of the red teaming process, reinforcing the need for community-driven learning. Offensive AI security is not merely about finding flaws; it's also about building stronger, more trustworthy AI systems. As AI becomes embedded in high-stakes domains such as healthcare, defense, and finance, its security posture will influence outcomes at scale. Continued collaboration between offensive and defensive teams will be essential to ensure these systems remain both capable and secure.

\section{Acknowledgments}
We would like to thank our colleague Dr. Mike Tan for the helpful conversations and insights that informed this work. His perspective contributed to our thinking during the development of the material, and we appreciate his input during the research process.

\bibliographystyle{IEEEtran}
\bibliography{references}

\end{document}